\documentclass[12pt,aps,epsfig,amssymb,tighten]{revtex4}
\usepackage{epsfig}
\begin{document}
\title{First-passage and extreme-value statistics of a particle\\
subject to a constant force plus a random force}
\author{
Theodore W. Burkhardt} \affiliation{Department of Physics, Temple University,
Philadelphia, PA 19122}
\date{\today}

\begin{abstract}
We consider a particle which moves on the $x$ axis and is subject to a constant
force, such as gravity, plus a random force in the form of Gaussian white noise. We
analyze the statistics of first arrival at point $x_1$ of a particle which starts at
$x_0$ with velocity $v_0$. The probability that the particle has not yet arrived at
$x_1$ after a time $t$, the mean time of first arrival, and the velocity
distribution at first arrival are all considered. We also study the statistics of
the first return of the particle to its starting point. Finally, we point out that
the extreme-value statistics of the particle and the first-passage statistics are
closely related, and we derive the distribution of the maximum displacement $m={\rm
max}_t[x(t)]$.
\end{abstract}
\maketitle \vskip 8cm \noindent email: tburk@temple.edu
 stochastic process, non-equilibrium statistics \clearpage

\section{Introduction}\label{intro}
\label{sec:intro} In this paper we consider a particle which moves on the $x$ axis
and is subject to both a constant force, such as gravity, and a random force in the
form of Gaussian white noise. The Newtonian equation of motion is given by
\begin{eqnarray}
&&\displaystyle{d^2x\over
dt^2}=g+\eta(t)\;,\label{eqmo}\\
&&\langle\eta(t)\rangle=0\;,\quad\langle \eta(t)\eta(t')\rangle=
2\Lambda\delta(t-t')\;,\label{randomforce}
\end{eqnarray}
where $g$ is a constant.

Simple stochastic processes such as this are of both mathematical and physical
interest. The approximately random collision forces experienced by a particular
particle in a many particle system are often modelled by Gaussian white noise.
Langevin's equation \cite{fpeq} for the motion of a Brownian particle in a constant
force field corresponds to Eq. (\ref{eqmo}) with an additional viscous damping term
of the form $-\lambda\thinspace dx/dt$ on the right-hand side. On setting $g=0$ in
Eq. (\ref{eqmo}) and regarding $t$ as a Cartesian coordinate instead of time, one
may interpret the path $x(t)$ of the particle as a configuration of a semi-flexible
polymer \cite{twb93}. For several applications of the process (\ref{eqmo}) related
to semi-flexible polymers and driven granular matter, see \cite{twb93,bk,twb07,ybg}
and references therein.

In this paper we study first-passage properties \cite{redner} of the process
(\ref{eqmo}). More precisely, we analyze the statistics of the first arrival at
point $x_1$ of a particle which starts at $x_0$ with velocity $v_0$. Due to
translational invariance no generality is lost in choosing $x_1$ to be the origin,
and since we consider both positive and negative $g$, no generality is lost in
choosing $x_0$ to be positive. Thus, ``first passage" corresponds to the first exit
of the particle from the positive $x$ axis. If the initial velocity $v_0$ is
positive, the particle must return to its initial position at least once before
exiting from the positive $x$ axis. Thus, in the limit $x_0\searrow 0$ with $v_0>0$,
the first-passage statistics reduces to the statistics of first return of the
particle to its initial position. Clearly, first-passage statistics, as defined
here, is the same as the statistics of absorption of a particle moving on the half
line $x=0$ with an absorbing boundary at $x=0$.

First-passage properties of the random acceleration process, corresponding to Eq.
(\ref{eqmo}) but without the constant term $g$ on the right hand side, are derived
or reviewed in Refs. \cite{mck,mw,sinai,twb93,twb07}. In the remainder of this
section we show how these results can be generalized to include the constant force.
In Section \ref{statquan} some statistical quantities of interest in connection with
first passage are defined, and in Section \ref{results} our explicit results are
presented. In Section \ref{extrstat} we show that the extreme-value statistics
\cite{twb07,gum,gal,ggetal,twbetal} and first-passage statistics of the process
(\ref{eqmo}) are closely related. This is then used in deriving the distribution of
the maximum displacement $m={\rm max}_t[x(t)]$ of a particle which begins at the
origin with velocity $v_0$.

With no loss of generality we replace the parameters $g$ and $\Lambda$, introduced
in Eqs. (\ref{eqmo}) and (\ref{randomforce}), by $g\to\gamma=\pm 1$ and $\Lambda=1$
throughout this paper, since this can be achieved by rescaling \cite{rescale} the
variables $x$ and $t$. For $\gamma=-1$ and $\gamma=1$, the constant force drives a
particle on the positive $x$ axis toward and away from the origin, respectively.

Integrating the equation of motion (\ref{eqmo}) yields
\begin{equation}
x(t)=x_0+v_0t+\textstyle{1\over 2}\displaystyle\thinspace\gamma t^2+\int_0^t
(t-t')\eta(t')dt'\label{x(t)}\;,
\end{equation}
which, together with properties (\ref{randomforce}) of the random force, implies the
moments
\begin{equation}
\langle x\rangle=x_0+v_0t+\textstyle{1\over 2}\displaystyle\gamma
t^2\;,\qquad\langle\left(x-\langle x\rangle\right)^2\rangle=\textstyle{1\over
3}\displaystyle\thinspace t^3\;.\label{moments}
\end{equation} Thus, the contribution of the random force on the right side of
Eq. (\ref{x(t)}) has typical size $t^{3/2}$. For large $t$ the constant force is
more important than the random force, but for small $t$ the opposite is true.

It is convenient to define $P_\gamma(x,v;x_0,v_0;t)\thinspace dxdv$ as the
probability that the position and velocity of a particle, moving according to Eq.
(\ref{eqmo}) with $g\to\gamma=\pm 1$ and $\Lambda=1$, evolve from $x_0$, $v_0$ to
values between $x$ and $x+dx$, $v$ and $v+dv$ in a time $t$ without ever reaching
$x=0$. The probability distribution $P_\gamma$ satisfies the time-dependent
Fokker-Planck equation \cite{fpeq}
\begin{equation} \left({\partial\over\partial t}+v{\partial\over\partial
x}+\gamma{\partial\over
\partial v}-{\partial^2\over
\partial v^2}\right)P_\gamma(x,v;x_0,v_0;t)=0\thinspace,\label{fp}
\end{equation}
with the initial condition
\begin{equation}
P_\gamma(x,v;x_0,v_0;0)=\delta(x-x_0)\delta(v-v_0)\;\label{ic} \end{equation} and
the boundary condition
\begin{equation}
P_\gamma(0,v;x_0,v_0;t)=0\;,\quad v>0\;.\label{bc}
\end{equation}
This boundary condition ensures that only trajectories which exit the positive $x$
axis for the first time at time $t$ contribute to $P_\gamma(0,v;x_0,v_0;t)$.
Trajectories which leave the positive $x$ axis at an earlier time and return to the
positive $x$ axis are excluded. Equation (\ref{bc}) is also the appropriate boundary
condition for motion on the half line $x>0$ with an absorbing boundary at $x=0$.

In the absence of the constant force, the corresponding probability distribution
$P_0(x,v;x_0,v_0;t)$ satisfies the same Fokker-Planck equation (\ref{fp}), initial
condition (\ref{ic}), and boundary condition (\ref{bc}), except that the term
$\gamma\partial P/\partial v$ in Eq. (\ref{fp}) is absent. This implies the relation
\begin{equation}
P_\gamma(x,v;x_0,v_0;t)=\exp\left[\textstyle{1\over
2}\gamma(v-v_0)-\textstyle{1\over 4}t\right]P_0(x,v;x_0,v_0;t)\label{Psigma}
\end{equation}
between the distributions with and without the constant force, which is central to
our work.

In a classic paper on the first-passage properties of a randomly accelerated
particle, McKean \cite{mck} derived the exact propagator $P_0(0,-v;0,v_0;t)$ for
$v>0$ and $v_0>0$, corresponding to a particle which leaves the origin with velocity
$v_0$ and returns for the first time at time $t$ with velocity $-v$ and speed $v$.
His result and Eq. (\ref{Psigma}) imply
\begin{eqnarray}
&&P_\gamma(0,-v;0,v_0;t)\nonumber\\
&&\quad\qquad={\sqrt{3}\over2\pi t^2}\thinspace \exp\left[-\textstyle{1\over
2}\thinspace\gamma(v+v_0)-\textstyle{1\over 4}\thinspace t-(v^2-vv_0+v_0^2)/t\right]
\thinspace{\rm erf}\left(\sqrt{3vv_0\over t}\right)\;,\label{propmck1}
\end{eqnarray} where ${\rm erf}(z)$ denotes the standard error function
\cite{as,gr}.

The Laplace transform
\begin{equation}
\tilde{P}_\gamma(x,v;x_0,v_0;s)=\int_0^\infty dt\thinspace
e^{-st}P_\gamma(x,v;x_0,v_0;t)\;.\label{laplacetransformdef}
\end{equation}
plays a central role in our work. Substituting Eq. (\ref{propmck1}) on the
right-hand side, using the integral representation \cite{as,gr} ${\ \rm
erf}(z)=2\pi^{-1/2}z\int_0^1dy\thinspace\exp(-z^2y^2)$, and integrating over $t$
with the help of Ref. \cite{gr}, we obtain
\begin{eqnarray}
&&\tilde{P}_\gamma(0,-v;0,v_0;s)\nonumber\\
&&\qquad={3\over 2\pi}\left(vv_0\right)^{1/2}e^{-\gamma(v+v_0)/2}\int_0^1
dy\thinspace
\exp\left[-(4s+1)^{1/2}(v^2-vv_0+v_0^2+3vv_0y^2)^{1/2}\right]\nonumber\\
&&\qquad\times\left[
(v^2-vv_0+v_0^2+3vv_0y^2)^{-3/2}+(4s+1)^{1/2}(v^2-vv_0+v_0^2+3vv_0y^2)^{-1}
\right].\label{laplacetransform1}
\end{eqnarray}
We will also need McKean's result \cite{mck} for the Laplace transform,
\begin{eqnarray}
&&\tilde{P}_\gamma(0,-v;0,v_0;s)\nonumber\\
&&\qquad={e^{-\gamma(v+v_0)/2}\over \pi^2 vv_0}\int_0^\infty
d\mu\thinspace\mu\thinspace{\sinh(\pi\mu)\over \cosh(\textstyle{1\over
3}\displaystyle\pi\mu)}\thinspace
K_{i\mu}(\sqrt{(4s+1)}\;v)K_{i\mu}(\sqrt{(4s+1)}\;v_0)\;, \label{laplacetransform2}
\end{eqnarray}
where $K_\mu(z)$ is a modified Bessel function \cite{as,gr}. Expressions
(\ref{laplacetransform1}) and (\ref{laplacetransform2}) are particularly convenient
for numerical and analytical calculations, respectively, and both expressions are
used below.

The exact solution $\tilde{P}_0(x,v;x_0,v_0;s)$ of the Fokker-Planck equation for
random acceleration on the half line $x>0$ with boundary condition (\ref{bc}), is
given in Ref. \cite{twb93}, where it is derived from more general results of
Marshall and Watson \cite{mw}. All of our results for first passage from an
arbitrary initial point $x_0$ are based on this solution. Substituting it in Eq.
(\ref{Psigma}) and setting $x=0$, we obtain
\begin{equation}
\tilde{P}_\gamma(0,-v;x_0,v_0;s)=e^{-\gamma(v+v_0)/2}\int_0^\infty dF e^{-F
x_0}\phi_{s+1/4,F}(-v)\psi_{s+1/4,F}(v_0) \;,\label{propmw}
\end{equation}
where
\begin{eqnarray}
\psi_{s,F}(v)&=&F^{-1/6}{\rm Ai}\left(F^{1/3}v+F^{-2/3}s\right)\;,\label{psi}\\
\phi_{s,F}(v)&=&\psi_{s,F}(v)-{1\over 2\pi}\int_0^\infty
dG\thinspace{\exp\left[-{2\over 3}s^{3/2}\left(F^{-1}+G^{-1}\right)\right]\over
F+G}\;\psi_{s,G}(-v)\;,\label{phi}
\end{eqnarray}
and ${\rm Ai}(z)$ is the Airy function \cite{as}. Some important properties of the
two set of basis functions $\psi_{s,F}(v)$ and $\phi_{s,F}(v)$ are discussed in Ref.
\cite{twb93}. For example, $\phi_{s,F}(v)$ vanishes identically for $v>0$, so that
Eq. (\ref{propmw}) satisfies the boundary condition (\ref{bc}).

\section{Statistical quantities of interest}\label{statquan}

The "survival probability" or probability that a particle with initial position and
velocity $x_0$ and $v_0$ has not yet left the positive $x$ axis after a time $t$ is
given by
\begin{equation}
 Q_\gamma(x_0,v_0;t)=\int_{-\infty}^\infty dv\int_0^\infty dx\thinspace
 P_\gamma(x,v;x_0,v_0;t)\;.
 \label{Qdef}
\end{equation}
According to Eqs. (\ref{fp}), (\ref{bc}), and (\ref{Qdef})
\begin{equation}
 {\partial\over\partial t}Q_\gamma(x_0,v_0;t)=
 -\int_0^\infty dv\;vP_\gamma(0,-v;x_0,v_0;t)\;.\label{probcons}
 \end{equation}
Thus, we interpret
\begin{equation}
vP_\gamma(0,-v;x_0,v_0;t)\thinspace dv\thinspace dt\;,\label{current}
\end{equation}
for $x_0>0$, as the probability that the particle reaches the origin for the first
time at a time between $t$ and $t+dt$ with speed between $v$ and $v+dv$.

Several useful relations follow from this interpretation of the quantity
(\ref{current}). The survival probability defined by Eq. (\ref{Qdef}), its limiting
value for $t\to\infty$, and its Laplace transform can be written in the form
\begin{eqnarray} &&Q_\gamma(x_0,v_0;t)=1-\int_0^t dt'\int_0^\infty dv\thinspace v
P_\gamma(0,-v;x_0,v_0;t')\;,\label{Q}\\
&&Q_\gamma(x_0,v_0;\infty)=1-\int_0^\infty dv\thinspace v
\tilde{P}_\gamma(0,-v;x_0,v_0;0)\;,\label{limQ}\\
&&\tilde{Q}_\gamma(x_0,v_0;s)={1\over s}\left[1-\int_0^\infty dv\thinspace v
\tilde{P}_\gamma(0,-v;x_0,v_0;s)\right]\;.\label{Qtilde}
\end{eqnarray}
The mean time to exit the positive $x$ axis for the first time is given by
\begin{eqnarray}
T_\gamma(x_0,v_0)&=&{\int_0^\infty dt\thinspace t\int_0^\infty dv\thinspace v
P_\gamma(0,-v;x_0,v_0;t)\over \int_0^\infty dt \int_0^\infty dv\thinspace v
P_\gamma(0,-v;x_0,v_0;t)}\label{T1}\\
&=&-\left[1-Q_\gamma(x_0,v_0;\infty)\right]^{-1}\thinspace\lim_{s\to
0}{\partial\over\partial s}\int_0^\infty dv\thinspace v
\tilde{P}_\gamma(0,-v;x_0,v_0;s)\label{T2}\;.
\end{eqnarray}

Since a particle which begins at the origin with a negative velocity immediately
moves onto the negative $x$ avis, $Q_\gamma$ and $T_\gamma$ satisfy the boundary
conditions
\begin{eqnarray}
&&Q_\gamma(0,v_0;t)=0\;,\quad v_0<0\;,\quad t>0\;,\label{Q0}\\
&&T_\gamma(0,v_0)=0\;,\quad v_0<0\;.\label{T0} \end{eqnarray}

Finally, the probability that the speed of the particle on exiting from the positive
$x$ axis for the first time is between $v$ and $v+dv$ is given by
$G_\gamma(v;x_0,v_0)\thinspace dv$, where
\begin{equation}
G_\gamma(v;x_0,v_0)={v\int_0^\infty dt\thinspace
P_\gamma(0,-v;x_0,v_0;t)\over\int_0^\infty dv\thinspace v\int_0^\infty dt\thinspace
P_\gamma(0,-v;x_0,v_0;t)}= {v\tilde{P}_\gamma(0,-v;x_0,v_0;0)\over
1-Q_\gamma(x_0,v_0;\infty)}\;,\label{velocitydist}\end{equation} and the
normalization $\int_0^\infty dv\thinspace G_\gamma(v;x_0,v_0)=1$ has been imposed.

\section{Results}\label{results}
\subsection{Limit $Q_\gamma(x_0,v_0;\infty)$ of the Survival
Probability}\label{subsec1} The survival probability $Q_\gamma(x_0,v_0;t)$
introduced in the preceding section can, in principle, be evaluated for arbitrary
$t$ from Eqs. (\ref{propmck1}), (\ref{propmw}), and (\ref{Q}). However, this
involves integrating over $t'$ and $v$ and, for $x_0>0$, inverting a Laplace
transform to go from $s$ to $t$, most of which must be performed numerically. In
this section we consider the limiting value $Q_\gamma(x_0,v_0;\infty)$ or
probability that in an infinite time the particle never leaves the positive $x$
axis, which can be obtained analytically.

In the absence of a constant force, $Q_0(x_0,v_0,t)$ decays as $t^{-1/4}$ in the
long-time limit \cite{mck,sinai,twb93}. Thus, $Q_0(x_0,v_0,\infty)=0$, which means
that the particle exits from the positive $x$ axis in an infinite time with
probability 1. As shown in Eq. (\ref{x(t)}), the constant force adds an extra term
${1\over 2}\thinspace\gamma t^2$ to the displacement $x(t)$ of the randomly
accelerated particle. In the case $\gamma=-1$ of a constant force toward the origin,
the particle leaves the positive $x$ axis sooner than without the constant force, so
\begin{equation}
Q_{-1}(x_0,v_0;\infty)=0\;.\label{Q1}
\end{equation}

In the case $\gamma=1$ of a constant force pushing the particle away from the
origin, the corresponding probability $Q_1(x_0,v_0,\infty)$ does not vanish and is
expected to increase as $x_0$ and $v_0$ increase. Recalling Eq. (\ref{Q0}),
substituting Eqs. (\ref{laplacetransform2}) and (\ref{propmw}) into Eq.
(\ref{limQ}), and proceeding as described in the Appendix, we obtain

\begin{equation}
Q_1(0,v_0;\infty)=\left\{\begin{array}{l}0\;,\quad v_0<0\;,\nonumber\\{\rm
erf}\left(\sqrt{\textstyle{3\over 2}\displaystyle\thinspace v_0}\right)\;,\quad
v_0>0\;,\end{array}\right.\label{Q2}
\end{equation}
\begin{eqnarray}
&&Q_1(x_0,v_0;\infty)=1-{e^{-v_0/2}\over\sqrt{2\pi}}\nonumber\\&&\qquad\qquad
\times\int_0^\infty dF\thinspace F^{-7/6}\exp\left(-\thinspace{1\over
12F}-Fx_0\right){\rm Ai}\left(F^{1/3}v_0+\textstyle{1\over 4}\displaystyle\thinspace
F^{-2/3}\right)\;.\label{Q3}
\end{eqnarray}
As in Eqs. (\ref{propmck1}) and (\ref{psi}), ${\rm erf}(z)$ denotes the error
function, and ${\rm Ai}(z)$ is the Airy function, both defined as in Ref. \cite{as}.
We found it useful to make the change of variables $F=u^{-6}$ in evaluating the
integrals in Eqs. (\ref{Q3}), (\ref{T4}), and (\ref{calP3}) numerically with {\em
Mathematica}.

Equations (\ref{Q2}) and (\ref{Q3}) imply the asymptotic behavior
\begin{eqnarray}
&&Q_1(0,v_0;\infty)\approx\left\{
\begin{array}{l}\displaystyle\left({6v_0\over\pi}\right)^{1/2}\;,\quad v_0\searrow
0\;,\\\displaystyle 1-\left({2\over 3\pi v_0}\right)^{1/2}e^{-3v_0/2}\;,\quad
v_0\to\infty\;,\end{array}\right.
\label{Qsmalllargev0}\\
&&Q_1(x_0,0;\infty)\approx\left\{
\begin{array}{l}\displaystyle{2^{5/6}3^{1/3}\over\Gamma\left({1\over 3}\right)}\thinspace
x_0^{1/6}\;,\quad x_0\searrow 0\;,\\\displaystyle 1-\left({3\over 8\pi^2
x_0}\right)^{1/4}e^{-(2x_0/3)^{1/2}}\;,\quad
x_0\to\infty\;,\end{array}\right.\label{Qsmalllargex0}\\
&&Q_1(x_0,v_0;\infty)\approx\left\{
\begin{array}{l}{\rm
erf}\left(\sqrt{\textstyle{3\over 2}\displaystyle\thinspace v_0}\right)
+\left[\displaystyle{\partial\over\partial
x}Q_1(x,v_0;\infty)\right]_{x=0}x_0\;,\quad
v_0>0\;,\quad x_0\searrow 0\;,\\
\displaystyle{3^{3/2}\over\sqrt{2\pi}}{x_0\over|v_0|^{5/2}}\thinspace
e^{-v_0/2-|v_0|^3/9x_0}\;,\quad v_0<0\;,\quad x_0\searrow 0\;,\\\displaystyle
1-\left({3\over 8\pi^2 x_0}\right)^{1/4}e^{-v_0-(2x_0/3)^{1/2}}\;,\quad
x_0\to\infty\;,\end{array}\right.
\end{eqnarray}

In Fig. 1, $Q_1(x_0,v_0;t)$, as given by Eqs. (\ref{Q2}) and (\ref{Q3}), is plotted
as a function of $v_0$ for several values of $x_0$. Note that $Q_1(x_0,v_0;t)$
increases monotonically with increasing $x_0$ and $v_0$, as expected.

If the random force is switched off, the particle trajectory becomes
$x(t)=x_0+v_0t+{1\over 2}\gamma t^2$, which never reaches the origin for $\gamma=1$
and $v_0>-\sqrt{2x_0}$. Thus, each of the smooth curves in Fig. 1 is replaced by the
unit step function $Q_1(x_0,v_0;\infty)=\theta(v_0+\sqrt{2x_0})$. The short vertical
lines in Fig. 1 indicate the value of $v_0$ at which $Q_1(x_0,v_0;\infty)$ is
discontinuous.

\subsection{Mean First-Passage Time $T_\gamma(x_0,v_0)$}\label{subsec2}
The $t^{-1/4}$ decay of the survival probability $Q_0(x_0,v_0,t)$ of a randomly
accelerated particle \cite{mck,sinai,twb93} is so slow that the mean time of its
first exit from the positive $x$ axis, given by $T_0(x_0,v_0)=\int_0^\infty
dt\thinspace t\left[-\partial Q_0(x_0,v_0,t)/\partial t\right]$, is infinite.
However, in the case $\gamma=-1$ of a constant force toward the origin in addition
to the random force, the corresponding mean time $T_{-1}(x_0,v_0)$ is finite.
Recalling Eq. (\ref{T0}), substituting Eqs. (\ref{laplacetransform2}),
(\ref{propmw}), (\ref{Q2}), and (\ref{Q3}) into Eq. (\ref{T2}), and proceeding as
described in the Appendix, we obtain
\begin{equation}
T_{-1}(0,v_0)=\left\{\begin{array}{l}0\;,\quad
v_0<0\;,\label{l}\\\displaystyle\sqrt{{2v_0\over\pi}}e^{-v_0/2}+\left({2\over
3}\displaystyle+v_0\right)\left[2-{\rm erfc}\left(\textstyle\sqrt{{1\over
2}\displaystyle v_0}\right)\right]\nonumber\\\qquad -\displaystyle{2\over
3}\displaystyle\;e^{3v_0/2}\thinspace{\rm erfc}\left(\sqrt{2v_0}\right)\;,\quad
v_0>0\;,\end{array}\right.\label{T3}
\end{equation}
\begin{eqnarray}
&&T_{-1}(x_0,v_0)={1\over 8}\sqrt{{3\over\pi}}\int_0^\infty dt\thinspace
t^{-3/2}\left(6x_0+2v_0t+t^2\right)\exp\left[-{3\over 4}\left({x_0+v_0t\over
t^{3/2}}-{1\over 2}\thinspace t^{1/2}\right)^2\right]\nonumber\\
&&\qquad\qquad -\;{e^{v_0/2}\over 2\pi}\int_0^\infty dF\thinspace
F^{-7/6}\exp\left(-\thinspace{1\over 12F}-Fx_0\right){\rm
Ai}\left(F^{1/3}v_0+\textstyle{1\over
4}\displaystyle\thinspace F^{-2/3}\right)\nonumber\\
&&\qquad\qquad\times\left[\sqrt{6\pi}-{\pi\over\sqrt{F}}\exp\left({1\over
6F}\right){\rm erfc}\left({1\over\sqrt{6F}}\right)\right]\;,\label{T4}
\end{eqnarray}
where ${\rm erfc}(z)=1-{\rm erf}(z)$ is the complementary error function \cite{as}.

Equations (\ref{T3}) and (\ref{T4}) imply the asymptotic behavior
\begin{eqnarray}
&&T_{-1}(0,v_0)\approx\left\{
\begin{array}{l}\displaystyle\left({18v_0\over\pi}\right)^{1/2}\;,\quad v_0\searrow
0\;,\\\displaystyle 2v_0+\textstyle{4\over 3}\displaystyle\;,\quad
v_0\to\infty\;,\end{array}\right.
\label{Tsmalllargev0}\\
&&T_{-1}(x_0,0)\approx\left\{
\begin{array}{l}\displaystyle\left({2\over\pi}\right)^{1/2}3^{5/6}\thinspace{\Gamma\left({5\over 6}\right)
\over\Gamma\left({2\over 3}\right)}\; x_0^{1/6}\;,\quad x_0\searrow
0\;,\\\displaystyle (2x_0)^{1/2}+\textstyle{2\over 3}\displaystyle\;,\quad
x_0\to\infty\;.\end{array}\right.\label{Tsmalllargex0}
\end{eqnarray}

The leading terms $2v_0$ and $(2x_0)^{1/2}$ in Eqs. (\ref{Tsmalllargev0}) and
(\ref{Tsmalllargex0}) for $x_0=0$, $v_0\to \infty$ and for $v_0=0$, $x_0\to\infty$
are easy to understand. According to the discussion following Eq. (\ref{moments}),
the constant force is more important than the random force for large $t$, implying
$x(t)\approx x_0+v_0t-{1\over 2}\thinspace t^2$, which vanishes at
$T_{-1}(x_0,v_0)\approx v_0+\sqrt{v_0^2+2x_0}$.

For small $x_0$ and $v_0$ the particle tends to reach the origin quickly, so
$T_\gamma(x_0,v_0)$ is small. For short times the random force is more important
than the constant force (see discussion following Eq. (\ref{moments})) and primarily
responsible for the asymptotic behavior $T_{\pm 1}(0,v_0)\sim v^{1/2}$ and
$T_{-1}(x_0,0)\sim x^{1/6}$ in Eqs. (\ref{Tsmalllargev0}), (\ref{Tsmalllargex0}),
and (\ref{Tplussmalllargev0}). We note that these same power laws for small $x_0$
and $v_0$ appear in the mean first exit time \cite{fr,mp} of a randomly accelerated
particle with initial position and velocity $x_0$, $v_0$ from the finite interval
$0<x<L$.

In Fig. 2, $T_{-1}(x_0,v_0)$, as given by Eqs. (\ref{T3}) and (\ref{T4}), is plotted
as a function of $v_0$ for several values of $x_0$. As expected, $T_{-1}(x_0,v_0;t)$
increases monotonically as $x_0$ and $v_0$ increase.

As discussed just above Eq. (\ref{Q2}), for $\gamma=1$ the probability
$Q_1(x_0,v_0;\infty)$ that the particle never leaves the positive $x$ axis is
nonzero, in general, and it increases, as in Eqs. (\ref{Qsmalllargev0}) and
(\ref{Qsmalllargex0}), as $v_0$ and $x_0$ increase. However, the mean first exit
time for those trajectories which do leave the positive $x$ axis, defined by Eq.
(\ref{T2}), is finite. From Eqs. (\ref{laplacetransform2}), (\ref{T2}), (\ref{T0}),
and (\ref{Q2}), we obtain
\begin{equation}
T_{1}(0,v_0)=\left\{\begin{array}{l}0\;,\quad
v_0<0\;,\nonumber\\\displaystyle{2\over 3}-v_0+\left(\sqrt{{6v_0\over\pi}}-{2\over
3}\right)\;{e^{-3v_0/2}\over{\rm erfc}\left(\sqrt{{3v_0\over 2}}\right)}\;,\quad
v_0>0\;,\end{array}\right.\label{T5}
\end{equation}
which has the asymptotic behavior
\begin{equation}T_1(0,v_0;\infty)\approx\left\{
\begin{array}{l}\displaystyle\left({2v_0\over 3\pi}\right)^{1/2}\;,\quad v_0\searrow
0\;,\\\displaystyle 2v_0-\left({2\pi v_0\over 3}\right)^{1/2}+{5\over 3}\;,\quad
v_0\to\infty\;.\end{array}\right. \label{Tplussmalllargev0}
\end{equation}

For arbitrary $x_0$ and $v_0$, $T_1(x_0,v_0)$ follows from substituting Eqs.
(\ref{propmw}) and (\ref{Q3}) into Eq. (\ref{T2}). This leads to a lengthy
expression, containing multiple integrals, which we were unable to simplify and omit
here.

The results (\ref{T3}) and (\ref{T5}) for $T_{-1}(0,v_0)$ and $T_1(0,v_0)$ are
compared in Fig. 2.

\subsection{Distribution $G_\gamma(v;x_0,v_0)$ of the Particle Speed at First Passage}
\label{subsec3}

For arbitrary initial position $x_0$ and initial velocity $v_0$, the distribution
$G_\gamma(v;x_0,v_0)$ of the particle speed on exiting from the positive $x$ axis
for the first time is determined by Eqs. (\ref{propmw}), (\ref{velocitydist}), and
(\ref{Q3}). Here we restrict our attention to the case of a particle which begins at
the origin with $v_0>0$. The distribution $G_\gamma(v;0,v_0)$ of its speed on
returning to the origin for the first time and leaving the positive $x$ axis can be
readily evaluated by substituting Eqs. (\ref{laplacetransform1}) and (\ref{Q2}) into
Eq. (\ref{velocitydist}) and integrating over the variable $y$ numerically. The
results, for several values of $v_0$, are shown in Fig. 3.

Each of the curves in Fig. 3 has a single peak. As $v_0$ increases, the peak shifts
to larger values of $v$, as expected, and becomes broader. The peak position and
width correspond roughly to the mean speed $\langle v\rangle_{_\gamma}$ at first
return and the root-mean-square deviation
$\sigma_{_\gamma}=\left\langle\left(v-\langle
v\rangle_{_\gamma}\right)^2\right\rangle_{_\gamma}^{1/2}$, where
\begin{equation}
\langle v^n\rangle_{_\gamma}= \int_0^\infty dv\thinspace v^n
G_\gamma(v;0,v_0)\;.\label{momv}
\end{equation}

Using Eqs. (\ref{laplacetransform2}), (\ref{velocitydist}), and (\ref{Q2}) and the
approach of the Appendix, we have calculated the first two moments of the speed at
first return analytically, obtaining
\begin{eqnarray}
&&\langle v\rangle_{_{-1}}=v_0+{4\over 3}\displaystyle +e^{-v_0/2}\nonumber\\
&&\quad\times\left[\left({2v_0\over\pi}\right)^{1/2}-\left(v_0+{2\over
3}\displaystyle\right)e^{v_0/2} {\rm erfc}\left(\sqrt{v_0\over 2}\right)-{2\over
3}\displaystyle\thinspace e^{2v_0}
{\rm erfc}\left(\sqrt{2v_0}\right)\right]\;,\label{momminus1}\\
&&\langle v^2\rangle_{_{-1}}=v_0^2+2v_0+2\langle v\rangle_{_{-1}}\;,\label{momminus2}\\
&&\langle v\rangle_{_{1}}={2\over 3}\left({e^{-3v_0/2}\over{\rm erfc}
\left(\sqrt{{3v_0\over 2}}\right)}-1\right)\;,\label{momplus1}\\
&&\langle v^2\rangle_{_{1}}=v_0^2-2v_0+{4\over 3}-\left[\left({2v_0\over
3\pi}\right)^{1/2}(v_0-5)+{4\over 3}\right]\left({3\over 2}\langle
v\rangle_{_{1}}+1\right)\;.\label{momplus2}
\end{eqnarray}
For large $v_0$ the average speed at first return and the root-mean-square deviation
from the average have the asymptotic forms
\begin{eqnarray}
&&\langle v\rangle_{_{-1}}\approx v_0+{4\over 3}\;,\label{momminus1large}\\
&&\sigma_{_{-1}} \approx \left({4\over 3}\thinspace
v_0\right)^{1/2}\;,\label{rmsminus}
\end{eqnarray}
and
\begin{eqnarray}
&&\langle v\rangle_{_1}\approx\left({2\pi v_0\over
3}\right)^{1/2}\;,\label{momplus1large}\\
&&\sigma_{_1}\approx \left[{1\over
3}\thinspace(8-2\pi)v_0\right]^{1/2}\;,\label{rmsplus}
\end{eqnarray}
which are qualitatively consistent with the evolution of the curves in Fig. 3 as
$v_0$ increases.

For large $v_0$ the constant force is more important than the random force, and
$x(t)\approx x_0+v_0t+\textstyle{1\over 2}\displaystyle\gamma t^2$. Thus, for
$\gamma=-1$, the particle returns to its starting point with approximately the same
speed it had initially. This is consistent with the asymptotic behavior in Eq.
(\ref{momminus1large}).

\section{Extreme-Value Statistics}\label{extrstat}
Consider a particle which begins at $x_0=0$ with velocity $v_0$ and moves according
to Eq.(\ref{eqmo}) with $g\to\gamma=\pm 1$. At some time in the interval
$0<t<\infty$ the particle attains a maximum displacement $m={\rm max}_t[x(t)]$. For
large $t$, $x(t)\approx{1\over 2}\gamma t^2$, as follows from the discussion below
Eq. (\ref{moments}). Thus, in the case $\gamma=1$ of a constant force in the
positive direction, $m=\infty$. In this section we consider the less trivial
question of the maximum displacement $m$ for $\gamma=-1$, and we derive the
corresponding distribution ${\cal P}_{-1}(m,v_0)$. Distributions such as this play a
central role in the field of extreme-value statistics \cite{twb07,gum,gal}. The
extreme-value statistics of a generalized Gaussian process that includes random
acceleration as a special case is studied in Refs. \cite{ggetal,twbetal}.

To derive the distribution ${\cal P}_{-1}(m,v_0)$, we begin by writing
\begin{equation}
{\cal P}_{-1}(m,v_0)={\partial\over\partial m}{\cal F}_{-1}1(m,v_0)\;,\label {calP1}
\end{equation}
where ${\cal F}_{-1}(m,v_0)$ is the probability that, for a constant force in the
{\em negative} direction, the displacement $x(t)$ of a particle which begins at the
origin with velocity $v_0$ never exceeds $m$ in the time interval $0<t<\infty$. For
$m<0$, ${\cal F}_{-1}(m,v_0)$=0, since the initial displacement $x_0=0$ already
exceeds $m$. For $m>0$, ${\cal F}_{-1}(m,v_0)$ is the same as the probability that,
for a constant force in the {\em positive} direction, a particle with initial
position $m$ and initial velocity $-v_0$ never reaches the origin. This follows from
the invariance of the probability under the coordinate transformation $x\to m-x$.
Since this latter probability is precisely the survival probability
$Q_1(m,-v_0;\infty)$ considered in Sections \ref{statquan} and \ref{results},
\begin{equation}
{\cal F}_{-1}(m,v_0)=\theta(m)\thinspace Q_1(m,-v_0;\infty)\;,\label{calF}
\end{equation}
where $\theta(m)$ is the standard step function.

Making use of Eqs. (\ref{calP1}) and (\ref{calF}) and the expressions for
$Q_1(0,v_0;\infty)$ and $Q_1(x_0,v_0;\infty)$ in Eqs. (\ref{Q2}) and (\ref{Q3}), we
obtain
\begin{eqnarray}
&&{\cal P}_{-1}(m,v_0)=\theta(-v_0)\thinspace{\rm erf}\left(\sqrt{\textstyle{3\over
2}\displaystyle\thinspace|v_0|}\right)\delta(m)\nonumber\\&&\quad
+\theta(m)\thinspace{e^{v_0/2}\over\sqrt{2\pi}}\int_0^\infty dF\thinspace
F^{-1/6}\exp\left(-\thinspace{1\over 12F}-Fm\right){\rm
Ai}\left(-F^{1/3}v_0+\textstyle{1\over 4}\displaystyle\thinspace
F^{-2/3}\right)\;\label{calP3}
\end{eqnarray}
for the extreme-value distribution. The distribution vanishes for $m<0$ and is
normalized so that $\int_{-\infty}^\infty dm\thinspace{\cal P}_{-1}(m,v_0)=1$, as
follows from Eqs. (\ref{calP1}) and (\ref{calF}) and the boundary condition
$Q_1(\infty,-v_0;\infty)=1$. The first term on the right side of Eq. (\ref{calP3})
has its origin in the non-zero probability ${\rm erf}\left(\sqrt{\textstyle{3\over
2}\displaystyle\thinspace|v_0|}\right)$ (see Section \ref{subsec1}) that a particle
which begins at the origin with $v_0<0$ never returns to the origin, in which case
the maximum displacement $m$ equals the initial value $x_0=0$.

The extreme-value distribution ${\cal P}_{-1}(m,v_0)$ is plotted as a function of
$m$ for several positive and negative values of $v_0$ in Figs. 4a and 4b,
respectively. In the absence of the random force, $x=v_0t-{1\over 2}t^2$, which
implies ${\cal P}_{-1}(m,v_0)=\theta(-v_0)\delta(m)+\theta(v_0)\delta(m-{1\over
2}v_0^2)$. The random force broadens the delta functions, as seen in the figure.

For positive $v_0$, the peak in Fig. 4a shifts to larger values of $m$ and becomes
broader as $v_0$ increases, as expected. The mean value and the root-mean-square
deviation vary as $\langle m\rangle\approx {1\over 2}v_0^2+v_0$ and
$\sigma\approx\left({2\over 3}v_0^3\right)^{1/2}$ for large positive $v_0$. For
large $v_0$ the constant force is more important than the random force, and the
leading term  in $\langle m\rangle$ equals the maximum displacement ${1\over
2}v_0^2$ of a particle subject only to the constant force.

In the results for $v_0<0$ in Fig. 4b, the vertical line at $m=0$ represents the
term proportional to $\delta(m)$ in Eq. (\ref{calP3}). The most probable value of
$m$, which maximizes ${\cal P}_{-1}(m,v_0)$, is zero for all negative $v_0$. The
mean value of $m$ is positive and, for $v_0$ negative and large in magnitude,
$\langle m\rangle\approx{4\over 3}({2|v_0|/3\pi })^{1/2}e^{-3|v_0|/2}$, and $\langle
m^2\rangle\approx{32\over 9}({2|v_0|^3/3\pi })^{1/2}e^{-3|v_0|/2}$.

\section{Concluding Remarks}\label{conclusion}
This completes our study of the first-passage and extreme-value statistics of the
process (\ref{eqmo}). In closing we note that in a mathematical tour de force,
Marshall and Watson \cite{mw} derived the Laplace transform
$\tilde{P}_{g,\lambda}(x,v;x_0,v_0;s)$ of the solution to the Klein-Kramers equation
with the absorbing boundary condition (\ref{bc}). The Klein-Kramers equation is the
Fokker-Planck equation for the process \cite{fpeq}
\begin{equation}
\displaystyle{d^2x\over dt^2}+\lambda{dx\over dt}=g+\eta(t)\;,\label{eqmo2}
\end{equation}
which, unlike Eq. (\ref{eqmo}), includes viscous damping and plays a central role in
the theory of Brownian motion. In principle, all of the first-passage and
extreme-value properties we have considered follow, for this more general process,
from the expression of Marshall and Watson for
$\tilde{P}_{g,\lambda}(x,v;x_0,v_0;s)$, which, however, involves an infinite double
sum over special functions and is difficult to work with.

\acknowledgements I am grateful to Zoltan R\'acz for asking about the extreme
statistics of the process (\ref{eqmo}), which led to its inclusion in this paper. It
is a pleasure to thank him and Dieter Forster for useful discussions and Robert
Intemann for help with {\it Mathematica} and LaTex.

\appendix
\section{Calculational Details}
The quantities $Q_\gamma(0,v_0;\infty)$ and $T_\gamma(0,v_0)$, defined in Eqs.
(\ref{limQ}) and (\ref{T2}), can both be expressed in terms of the integral
\begin{eqnarray}
&&\int_0^\infty dv\thinspace v
\tilde{P}_\gamma(0,-v;0,v_0;s)\nonumber\\
\qquad\qquad&&={e^{-\gamma v_0/2}\over \pi^2 v_0}\int_0^\infty
d\mu\thinspace\mu\thinspace{\sinh(\pi\mu)\over \cosh({1\over 3}\pi\mu)}\thinspace
F_\gamma(\mu,s)K_{i\mu}(\sqrt{(4s+1)}\;v_0)\;,\label{basiceq1}\\
&&\qquad\qquad F_\gamma(\mu,s)=\int_0^\infty dv\thinspace e^{-\gamma
v/2}K_{i\mu}(\sqrt{(4s+1)}\;v)\;,\label{Fgammamu}
\end{eqnarray}
where we have used the expression for $\tilde{P}_\gamma(0,-v;0,v_0;s)$ in Eq.
(\ref{laplacetransform2}). Evaluating $F_\gamma(\mu,s)$ with the help of the
integral representation \cite{as,gr}
\begin{equation}
K_{i\mu}(v)=\int_0^\infty dt\thinspace\cos(\mu t)\thinspace e^{-v\cosh
t}\label{intrep}
\end{equation}
and substituting the result in Eq. (\ref{basiceq1}), we obtain
\begin{eqnarray}
&&\int_0^\infty dv\thinspace v \tilde{P}_\gamma(0,-v;x_0,v_0;s)={e^{-\gamma
v_0/2}\over\pi v_0}\textstyle{\left(4s+1-{1\over
4}\gamma^2\right)^{-1/2}}\nonumber\\
&&\qquad\displaystyle\times\int_0^\infty
d\mu\thinspace\mu\thinspace{\sinh\left[\mu\arccos\left({1\over
2}\gamma(4s+1)^{-1/2}\right)\right]\over\cosh\left({1\over
3}\pi\mu\right)}\thinspace K_{i\mu}(\sqrt{(4s+1)}\;v_0)\;.\label{basiceq2}
\end{eqnarray}

First we consider the quantity $Q_\gamma(0,v_0;\infty)$, defined in Eq.
(\ref{limQ}). For $\gamma=-1$ and $s=0$, the $\arccos$ in Eq. (\ref{basiceq2})
equals ${2\over 3}\pi$. From Eqs. (\ref{limQ}) and (\ref{basiceq2}) and the relation
\cite{gr}
\begin{equation}
\int_0^\infty d\mu\thinspace\mu\thinspace\sinh(b\mu)K_{i\mu}(v_0)={1\over
2}\thinspace\pi v_0\sin b\thinspace e^{-v_0\cos b}\;,\label{integ1}
\end{equation}
we obtain $Q_{-1}(0,v_0;\infty)=0$ for $v_0>0$, in agreement with Eq. (\ref{Q1}).

For $\gamma=1$ and $s=0$, the $\arccos$ in Eq. (\ref{basiceq2}) equals ${1\over
3}\pi$, and expression (\ref{Q2}) for $Q_1(0,v_0;\infty)$, with $v_0>0$, follows
from Eqs. (\ref{limQ}) and (\ref{basiceq2}) and the relation
\begin{equation}
\int_0^\infty d\mu\thinspace\mu\thinspace\tanh\left(\textstyle{1\over
3}\displaystyle\pi\mu\right)K_{i\mu}(v_0)={\sqrt{3}\over 2}\thinspace\pi v_0
e^{v_0/2}{\rm erfc}\left(\sqrt{{3v_0\over 2}}\right)\;,\label{integ2}
\end{equation}
which we derived with the help of the integral representation (\ref{intrep}).

We now turn to $Q_\gamma(x_0,v_0;\infty)$ for $x_0\neq 0$. The result
$Q_{-1}(x_0,v_0;\infty)=0$ was established in the paragraph containing Eq.
(\ref{Q1}). Expression (\ref{Q3}) for $Q_1(x_0,v_0;\infty)$ follows from Eqs.
(\ref{propmw})-(\ref{phi}) and (\ref{limQ}). The lengthy derivation will not be
given here, but it is easy to see, with the help of  Ref. \cite{twb93}, that the
result (\ref{Q3}) satisfies the appropriate Fokker-Planck equation
$(v_0\partial_{x_0}+\gamma\partial_{v_0}+
\partial_{v_0}^2)Q_\gamma(x_0,v_0;\infty)=0$ and to check, by numerical integration, that
Eqs. (\ref{Q2}) and (\ref{Q3}) agree for $x_0=0$.

The results (\ref{T3}) and (\ref{T5}) for $T_\gamma(0,v_0)$ may be derived by
substituting Eqs. (\ref{Q2}) and (\ref{basiceq2}) into Eq. (\ref{T2}) and using Eqs.
(\ref{integ1}), (\ref{integ2}) and some analogous integrals over $\mu$ which can be
evaluated with the help of the integral representation (\ref{intrep}). Expression
(\ref{T4}) for $T_{-1}(x_0,v_0)$ follows from substituting Eqs.
(\ref{propmw})-(\ref{phi}) and (\ref{Q1}) into Eq. (\ref{T2}), but the derivation is
long and will not be given here. Making use of Ref. \cite{twb93}, we have confirmed
that the result (\ref{T4}) satisfies the appropriate Fokker-Planck equation
$(v_0\partial_{x_0}+\gamma\partial_{v_0}+\partial_{v_0}^2)T_\gamma(x_0,v_0;\infty)=-1$
and checked by numerical integration that Eqs. (\ref{T3}) and (\ref{T4}) agree for
$x_0=0$.

The moments $\langle v^n\rangle_{_\gamma}\thinspace,$ defined by Eqs.
(\ref{velocitydist}) and (\ref{momv}), can be expressed in terms in terms of the
integral
\begin{eqnarray}
&&\int_0^\infty dv\thinspace v^{n+1} \tilde{P}_\gamma(0,-v;x_0,v_0;s)={e^{-\gamma
v_0/2}\over\pi
v_0}\left(-2{\partial\over\partial\gamma}\right)^n\left\{\textstyle\left(4s+1-{1\over
4}\gamma^2\right)^{-1/2}\right.\nonumber\\&&\qquad\left.\times\displaystyle\int_0^\infty
d\mu\thinspace\mu\thinspace{\sinh\left[\mu\arccos\left({1\over
2}\gamma(4s+1)^{-1/2}\right)\right]\over\cosh\left({1\over 3
}\pi\mu\right)}\thinspace K_{i\mu}(\sqrt{(4s+1)}\;v_0)\right\}\;.\label{basiceq3}
\end{eqnarray}
This relation is the same as Eq. (\ref{basiceq2}) except for the extra factor $v^n$
introduced by applying $(-2\partial/\partial\gamma)^n$ to the quantity
$F_\gamma(\mu,s)$ in Eq. (\ref{Fgammamu}). The steps leading from Eq.
(\ref{basiceq3}) to the results for $\langle v\rangle_{_\gamma}$ and $\langle
v^2\rangle_{_\gamma}$ in Eqs. (\ref{momminus1})-(\ref{momplus2}) are very similar to
the steps, described above, from Eq. (\ref{basiceq2}) to the final expressions for
$Q_\gamma(0,v_0;\infty)$ and $T_\gamma(0,v_0)$.


\clearpage
\begin{figure}[fig1]
\begin{center}
   \includegraphics[width=12cm]{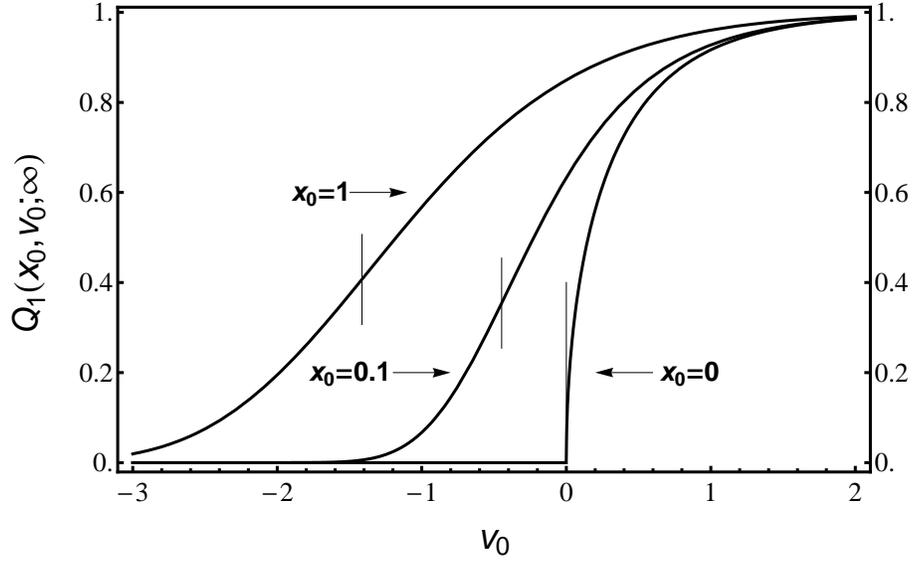}
\end{center}
\caption{Probability $Q_1(x_0,v_0;\infty)$, given by Eqs. (\ref{Q2}) and (\ref{Q3}),
that a particle subject to a random force plus a constant force pushing it away from
the origin never leaves the positive $x$ axis. If the random force is switched off,
$Q_1(x_0,v_0;\infty)$ becomes a unit step function, as discussed in the second
paragraph below Eq. (\ref{Qsmalllargex0}). The short vertical lines in the figure
indicate the values of $v_0$ at which the step function jumps from 0 to 1 for
$x_0=0$, 0.1, 1.}\label{fig1}
\end{figure}

\clearpage
\begin{figure}[fig2]
\begin{center}
   \includegraphics[width=12cm]{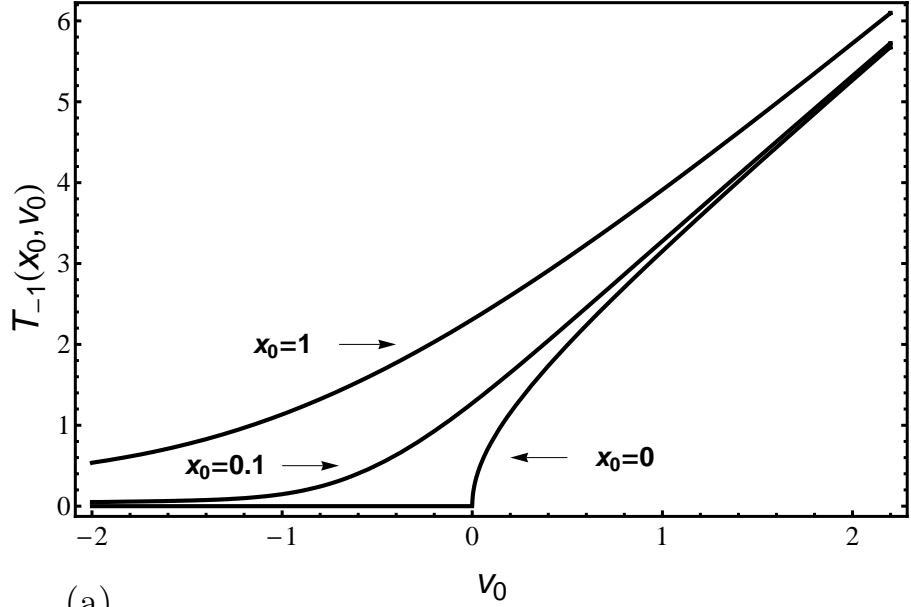}\vskip 1.5cm
   \end{center}
   \vspace{-2.75cm}
   \begin{flushleft}
   \hspace{3cm}\large{(a)}\\
   \end{flushleft}
   \vspace{0.5cm}
   \begin{center}
   \includegraphics[width=12cm]{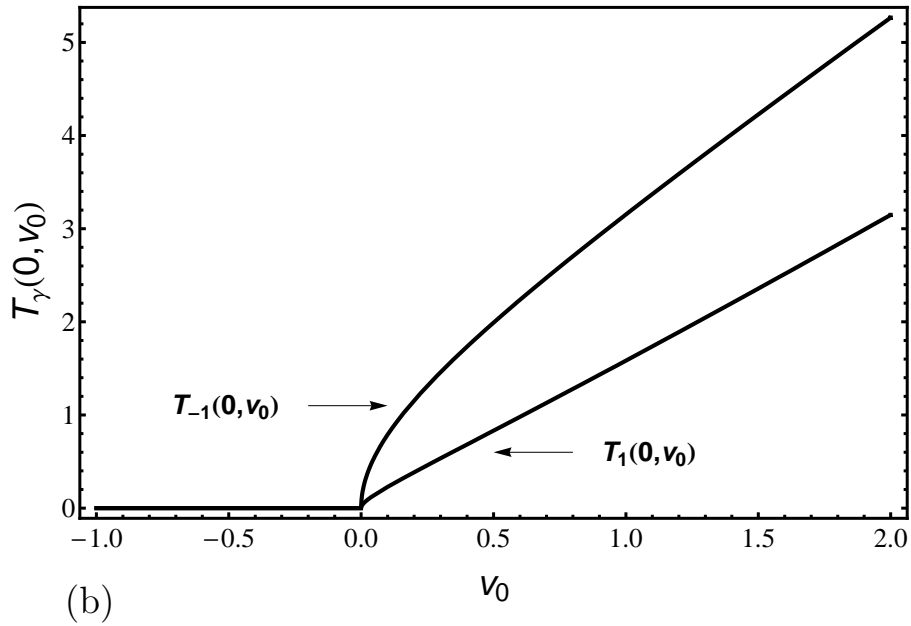}
\end{center}
   \vspace{-1.75cm}
   \begin{flushleft}
   \hspace{3cm}\large{(b)}\\
   \end{flushleft}
\caption{(a) Mean time $T_{-1}(x_0,v_0)$ to exit the positive $x$ axis for the first
time for a particle subject to a random force plus a constant force toward the
origin, given by Eqs. (\ref{T3}) and (\ref{T4}). (b) Mean first exit times
$T_{-1}(0,v_0)$ and $T_1(0,v_0)$, given in Eqs. (\ref{T3}) and (\ref{T5}) for
constant forces toward and away from the origin, respectively. }\label{fig2}
\end{figure}

\clearpage
\begin{figure}[fig3]
\begin{center}
   \includegraphics[width=12cm]{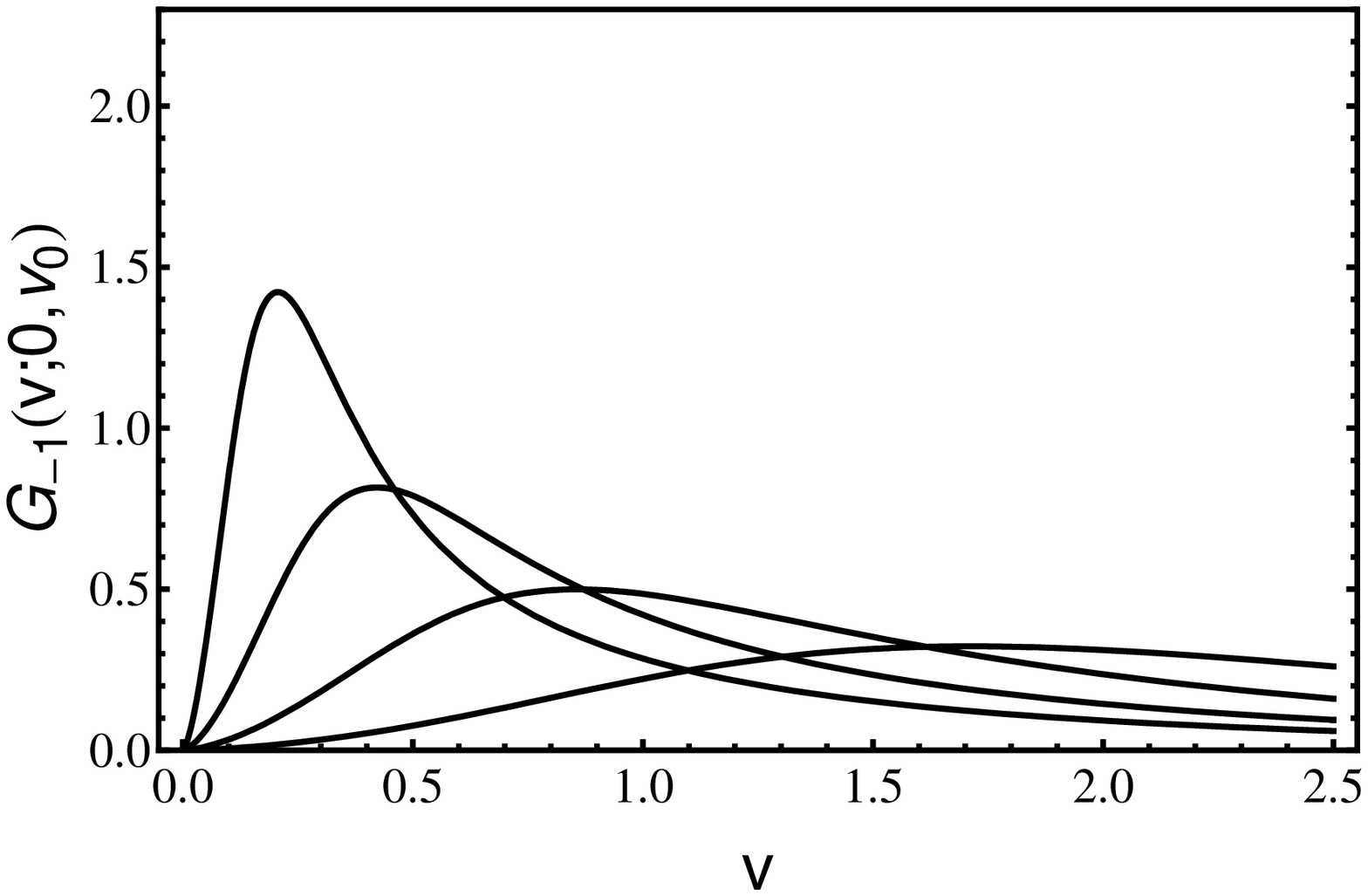}\vskip 1.5cm
   \end{center}
   \vspace{-2.75cm}
   \begin{flushleft}
   \hspace{3cm}\large{(a)}\\
   \end{flushleft}
   \vspace{0.5cm}
   \begin{center}
   \includegraphics[width=12cm]{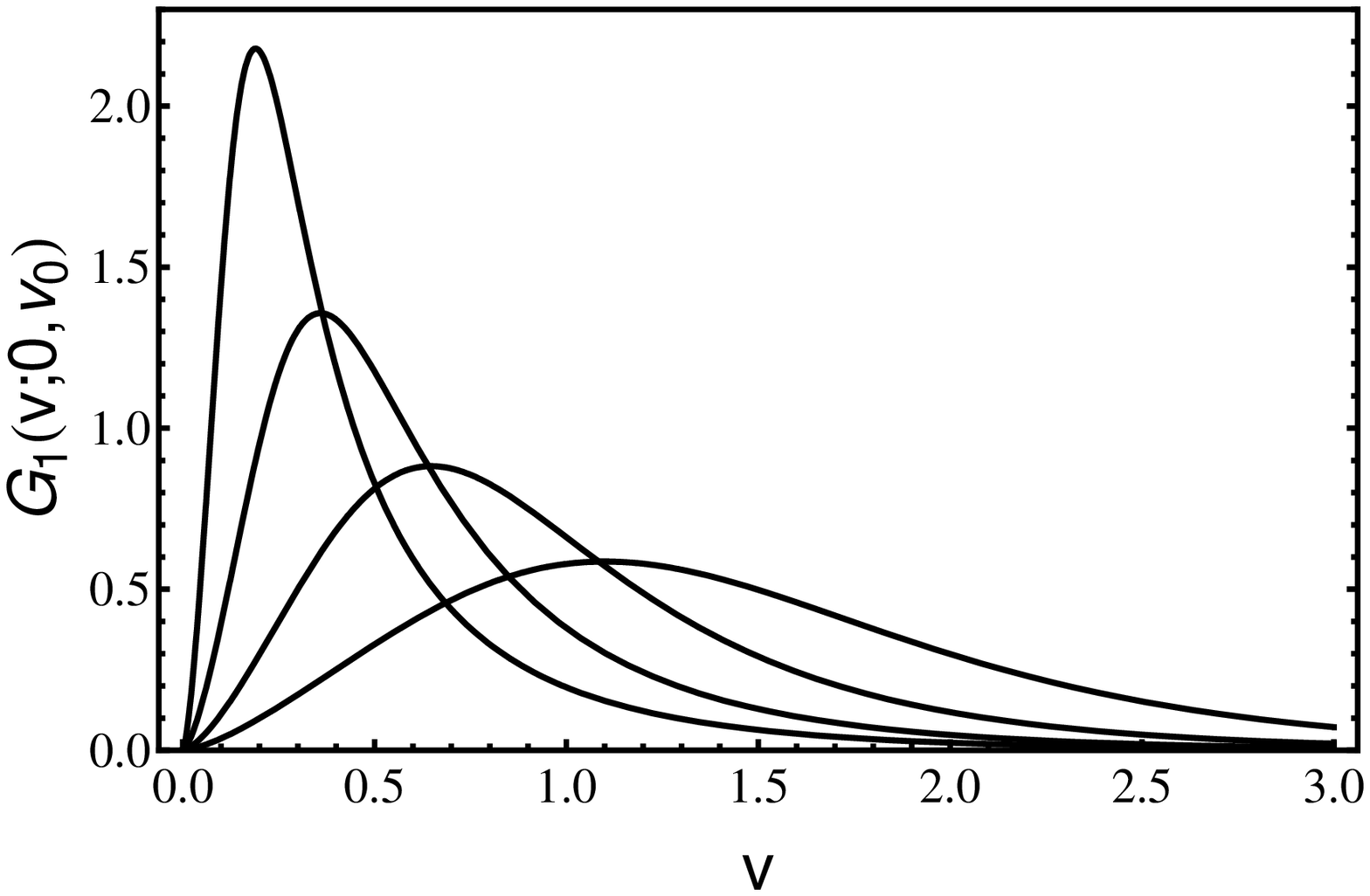}
\end{center}
   \vspace{-1.75cm}
   \begin{flushleft}
   \hspace{3cm}\large{(b)}\\
   \end{flushleft}
\caption{Distribution $G_\gamma(v;0,v_0)$, given by Eqs. (\ref{laplacetransform1}),
(\ref{velocitydist}), and (\ref{Q2}), of the particle speed $v$ at first return, for
$v_0=$ 0.2, 0.4, 0.8, and 1.6. The results in (a) and (b) are for constant forces
directed toward and away from the origin, respectively. As $v_0$ increases, the peak
becomes lower and broader and moves to the right.}\label{fig3}
\end{figure}
\clearpage

\clearpage
\begin{figure}[fig4]
\begin{center}
   \includegraphics[width=12cm]{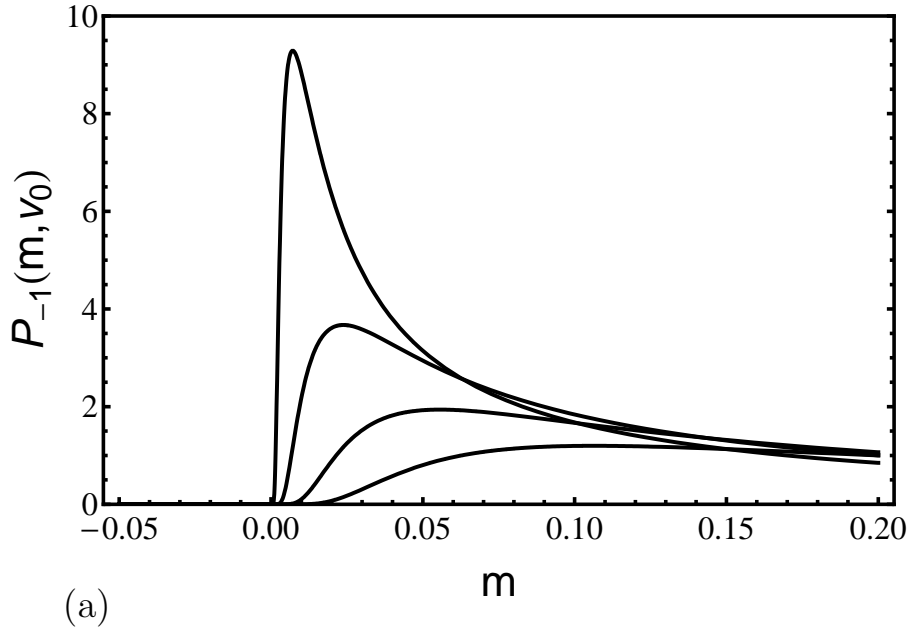}\vskip 1.5cm
   \end{center}
   \vspace{-2.75cm}
   \begin{flushleft}
   \hspace{3cm}\large{(a)}\\
   \end{flushleft}
   \vspace{0.5cm}
   \begin{center}
   \includegraphics[width=12cm]{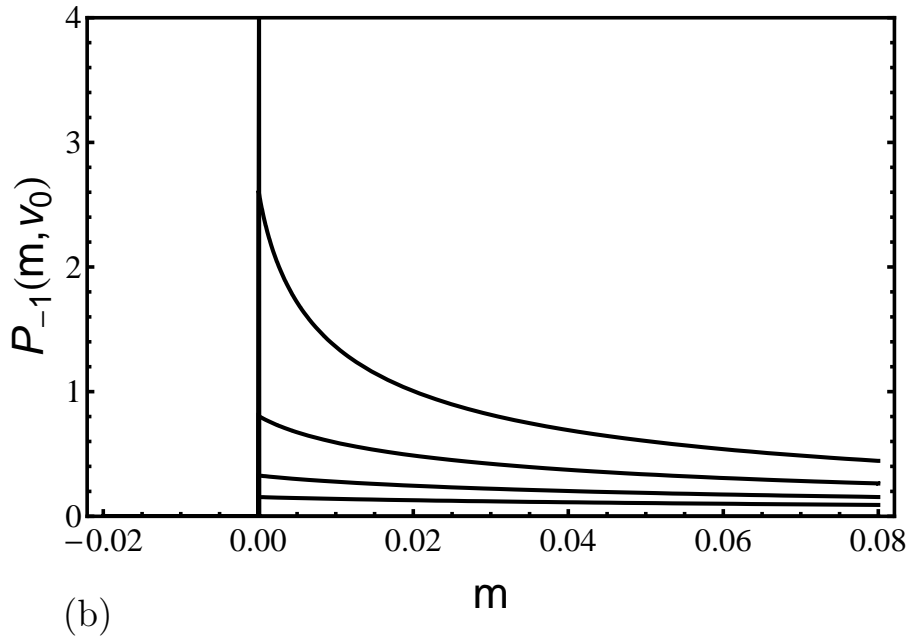}
\end{center}
   \vspace{-1.75cm}
   \begin{flushleft}
   \hspace{3cm}\large{(b)}\\
   \end{flushleft}
\caption{Distribution ${\cal P}_{-1}(m,v_0)$, given by Eq. (\ref{calP3}), of the
maximum displacement $m$ attained by a particle which begins at the origin with
velocity $v_0$ and moves according to Eq. (\ref{eqmo}) with $g=-1$. The results in
(a) are for $v_0=$ 0.4, 0.6, 0.8, and 1.0. As $v_0$ increases, the peak becomes
lower and broader and moves to the right. The curves in (b) correspond, from top to
bottom, to $v_0=$ -0.4, -0.6, -0.8, and -1. The vertical line at $m=0$ represents
the term in $P_{-1}(m,v_0)$ proportional to $\delta(m)$.}\label{fig3}
\end{figure}
\end{document}